\documentclass[aps,pre,reprint]{revtex4-1} 

\usepackage[dvips]{graphicx}
\usepackage{epsfig}
\usepackage{amsmath}
\usepackage{color}
\usepackage{hyperref}

\begin{document}

\title{Nucleation of a new phase on a surface that is changing irreversibly with time}

\author{Richard P. Sear}
\email{r.sear@surrey.ac.uk}

\affiliation{Department of Physics, University of Surrey\\ Guildford, Surrey
GU2 7XH, United Kingdom}

\begin{abstract}
Nucleation of a new phase almost
always starts at a surface.
This surface is almost always assumed
not to change with time. However, surfaces can roughen, partially dissolve and
change chemically with time. Each of these irreversible changes will change
the nucleation rate at the surface, resulting in a time-dependent nucleation rate.
Here we
use a simple model to show that partial surface dissolution can qualitatively
change the nucleation process, in a way that is testable in experiment.
The changing surface means that the nucleation rate is increasing
with time. There is an initial period during which no nucleation
occurs, followed by relatively rapid nucleation.
\end{abstract}

\maketitle 

\section{Introduction}

Nucleation of a new phase, such as a crystal or a liquid, almost
always occurs with the nucleus
at a surface \cite{debenedetti,sear_rev,sear_rev12}.
The rate of nucleation
is known to be extremely sensitive to microscopic details
of the surface.
For example, microscopic changes in the surface geometry
\cite{page06,page09_jacs,sear_rev12,hedges12,meel10}
can change rates by orders of magnitude.
It is also well known that surfaces change irreversibly over time.
Mineral surfaces in the atmosphere are subject to weathering \cite{white03},
smooth metal surfaces corrode and become pitted \cite{callister_book}, and
polymer surfaces degrade \cite{callister_book}.
Despite the fact that forming a nanoscale pit is known to hugely increase
the nucleation rate \cite{page06,meel10,hedges12}, no attempts
have been made to model and understand the effect on nucleation, of pitting and
other dynamic changes in a surface.
We model this here, using a simple generic model of nucleation of
a new phase on a slowly
dissolving
surface. In this simple model, we find that dissolution,
by roughening the surface, greatly increases
the nucleation rate, i.e., the nucleation rate, $r_N(t)$, becomes
an increasing function of time.

The classical picture of nucleation of a new crystal is 
that the nucleus is a large and rare
thermal fluctuation of the nucleating phase. The rate is low due to the rareness
of the fluctuation; it is necessary to wait a long time for such a rare fluctuation
to occur. In this classical picture of nucleation,
the fluctuation
occurs in a system that is at a metastable local equilibrium
and so is not changing with time
\cite{debenedetti,sear_rev,sear_rev12}.
Nucleation is then a stochastic process, with a time-independent
nucleation rate, $r_N$ (that depends exponentially
on the free energy cost of the fluctuation).
For a stochastic process with a time-independent rate,
the probability that nucleation has {\em not}
occurred at a time $t$, is an exponential function
of time: $P(t)=\exp(-r_Nt)$.
However, if the surface is changing with time, so will the
free energy of forming a nucleus and hence the nucleation rate,
and $P(t)$ is no longer an exponential function
of time.

The property $P(t)$ is easily measurable in experiment, and this has been
done for a number of crystallising systems
\cite{duft04,carvalho11,diao11_langmuir,diao12_cgd,murray11,kulkarni13,toldy12,kim13}.
For crystallisation,
$P(t)$ can be estimated in experiment by preparing
a set of tens or more identical droplets and then plotting the fraction of them
in which crystals have nucleated, as a function of time.
The signature of a rate increasing with time, 
is an initial plateau with $P(t)\simeq 1$ due to a small initial nucleation rate,
followed by $P(t)$ dropping faster and faster, and so curving downwards, as the
nucleation rate increases. 
This signature has been in observed in
the crystallisation of the explosive RDX by Kim {\it et al.}~\cite{kim13}.
Thus our model provides a possible qualitative explanation for this observation.
It is only qualitative as our model is a simple lattice model, not an accurate
model of RDX in solution, and the experiments do not characterise the
surface that nucleation is occurring on. Further experiments to characterise the
source of the increasing nucleation rate, and simulation of more detailed
models will be needed to develop our understanding
of time-dependent nucleation rates.

In the next section we will describe our simple model, and the
simulation algorithm we will use. We will then present and discuss results,
before comparing these results with a simple analytic theory.
We discuss the generality of our observed $P(t)$ in nucleation and note
that a similar form is seen for cancer. The final section is a conclusion.

\begin{figure}[bt!]
\includegraphics[width=8.0cm]{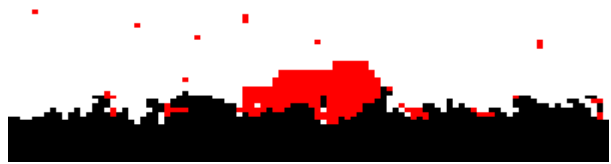}
\caption{(Color online) Snapshot of nucleation of a new phase (red) in a solution (white)
in contact with a slowly dissolving surface (black).
The snapshot is of a simulation in a box of 100 ($x$ axis) by
400 ($y$ axis) lattice
sites, but it is cropped along the vertical axis.
The dissolution rate $r_D=2\times 10^{-6}$, $2h/kT=0.12$, and
the snapshot is taken with the nucleus a little over the
nucleation barrier.
}
\label{fig:snapshot}
\end{figure}

\section{Model}

Our simple model for nucleation on a
slowly dissolving surface
is a modified two-dimensional Ising model or lattice gas
(the two models are equivalent \cite{chandler_book}), 
with Glauber Monte Carlo dynamics \cite{binder_book10}. See
Fig.~\ref{fig:snapshot} for a simulation snapshot.
We will
use lattice-gas terminology here.
Each lattice site is either empty (left white in Fig.~\ref{fig:snapshot}),
filled with a particle (red) or part of the
surface and so filled by a surface particle (black).
We will study
nucleation of the high-particle-density (liquid so mostly
red) phase from the dilute phase (vapour so mostly white),
as we did in
earlier work on nucleation in pores \cite{page06}.
Nucleation is via a rare barrier-crossing thermal fluctuation,
that occurs on an irreversibly changing surface.

\begin{figure}[tb!]
\vspace*{0.5cm}
\includegraphics[width=6.0cm]{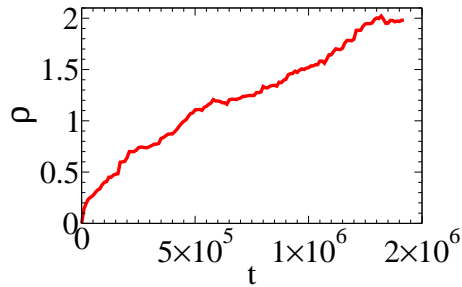}\\
\caption{(Color online) Plot of the surface roughness, $\rho$, for a single surface,
as a function of time $t$ in units of cycles. The run is until
nucleation occurs.
The dissolution rate $r_D=2\times 10^{-6}$.
}
\label{fig:rough}
\end{figure}

Neighbouring particles interact via an energy $\epsilon$,
and a particle interacts with a neighbouring surface particle
with an energy $\epsilon/2$. This results in the nucleus of the liquid phase
having a contact angle $\theta=90^{\circ}$ with the surface.
We work at a temperature such that $\epsilon/kT=3.0$.
The chemical potential of the particles $\mu/kT=2h/kT-2\epsilon/kT$,
where $2h/kT$ is the supersaturation in units of $kT$ ($h$ is the magnetic
field in spin language).
All our results are for simulations of systems of 100 ($x$)
by 400 ($y$) lattice sites.
Our unit of time is one cycle: one attempted Monte Carlo move per site.

Each surface starts out as perfectly flat, and
parallel to the $x$ axis, but dissolves at a rate
$r_D$ along $y$ during the simulation.
Dissolution
and nucleation is shown in Supplementary Movie 1 \cite{supp_mat_pre14}.
Our model for dissolution is simple.
In our Monte Carlo simulations
we select lattice sites at random. If a site is occupied by a surface
particle,
then if all 4 of its neighbours are also surface particles we do nothing.
If between 1 and 3 of its neighbours are surface particles, we flip
it irreversibly to an empty site with probability $r_D$. If none of its
4 neighbours are surface particles we flip it irreversibly to an empty
site with probability 1. This is a simple model of irreversible dissolution
that creates rough surfaces.
We measure height variations in the surface
by $\rho$, the standard deviation
of the surface height. The height is defined as the $y$ coordinate
of the highest lattice site occupied by a surface particle.
The roughness as a function of time is plotted
for a single run in Fig.~\ref{fig:rough}.

To obtain statistics of the nucleation times we simply run many
simulations, each time starting with a perfectly smooth surface.
Nucleation is defined as having occurred when the
fraction of particles is greater than 10\%. This fraction
is defined as being the ratio of the number of particles
to the total number of particles and empty sites.
The equilibrium fraction of sites
occupied by particles in the dilute phase is $\simeq 0.5$\%, so
this threshold is only crossed once nucleation has occurred.

\section{Computer Simulation Results}

\begin{figure}[tb!]
\includegraphics[width=7.5cm]{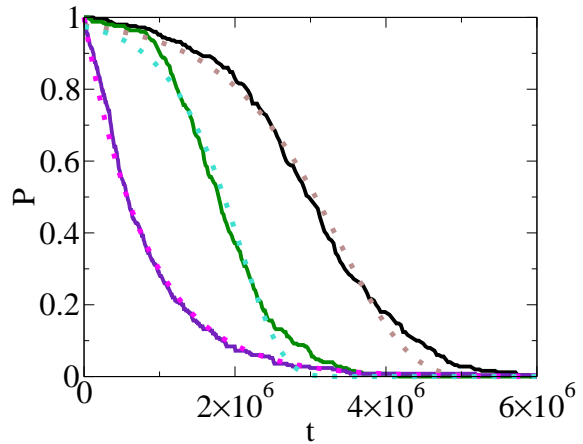}\\
\caption{(Color online) Plot of the cumulative probability that nucleation has not
occurred, $P(t)$, as a function of time, $t$, in units of cycles.
The black (right) and green (middle) curves are with dissolution, at rates
$r_D=10^{-6}$ and $r_D=2\times 10^{-6}$, respectively.
In both cases, the supersaturation $2h/kT=0.12$.
The purple (left) curve is at a higher supersaturation 
$2h/kT=0.16$, and with no dissolution ($r_D=0$).
The magenta dotted curve is a fit of an exponential function
to the $P(t)$ without dissolution.
The brown and turquoise
dotted curves are fits of the function of Eq.~(\ref{eq:exp_p})
to the black and green $P(t)$'s respectively.
In all cases the simulation $P(t)$'s are obtained from 250 nucleation runs.
}
\label{fig:bigp}
\end{figure}

Results for the probability that nucleation has
not occurred, $P(t)$, are plotted in Fig.~\ref{fig:bigp}.
In Fig.~\ref{fig:bigp} we plot $P(t)$ for systems with
and without surface dissolution. The purple curve
is for no dissolution, and at a higher supersaturation.
We see that this $P(t)$ is well fit by a simple exponential function.
The fitted rate is
$1.22\times 10^{-6}$. This agrees well with
the nucleation rate of
$1.35\times 10^{-6}\pm 0.22\times 10^{-6}$ for this surface calculated
using the Forward Flux Sampling (FFS)
method of Allen {\it et al.}~\cite{allen05,allen09,page06}.
This is just what we expect for a time-independent
nucleation rate. Note that the
curve without dissolution
in Fig.~\ref{fig:bigp} cannot be directly compared with the curves
for the surfaces with dissolution, because the supersaturations are
different.

However, with slow dissolution (black and green curves)
the functional form is very
different. In particular, whereas the exponential function is steepest
at $t=0$, with dissolution the $P(t)$ is almost
horizontal there. Initially, there is a waiting period
during which almost no nucleation occurs and so $P(t)$ is
almost horizontal, then
significant nucleation occurs and $P(t)$ drops.
The waiting time is increased if the dissolution rate is
decreased.
The black curve is for a $r_D$ half that of the green curve,
and we see that the initial plateau at short times is longer.

The form of $P(t)$ with dissolution
is also very different from the stretched exponential $P(t)$
that results from quenched (i.e., time-independent) disorder
\cite{sear12,sear13}. Thus the shape of $P(t)$ allows us to distinguish
between systems with a single time-independent rate (simple exponential),
systems with quenched disorder (stretched exponential),
and systems with a time-dependent nucleation rate due to
dissolution (initially almost horizontal $P(t)$).
See Refs.~\onlinecite{duft04,carvalho11} for
examples of experimental systems with simple exponential $P(t)$,
and Ref.~\onlinecite{diao12} for an experimental system showing
stretched exponential behaviour.
I am
aware of two experimental studies, which show $P(t)$'s with a plateau
at short times. Both are on crystallisation from solution,
of glycine in the case of Badruddoza {\it et al.}~\cite{badruddoza13},
and of the explosive RDX in the case of the work of
Kim {\it et al.}~\cite{kim13}. In both sets of results,
the effective nucleation rate is clearly increasing with time,
and so they
are both consistent with nucleation occurring on a surface
that is changing with time in such a way as to increase the nucleation rate.
Determining if a time-dependent surface is responsible
for the form of the experimental $P(t)$,
would require finding the surface
that nucleation is occurring on. This was not done in these experiments
but could be attempted in future work.
As estimating $P(t)$ is straightforward in experiment
\cite{diao11_langmuir,diao12_cgd,murray11,kulkarni13},
$P(t)$ is probably the best way to obtain
evidence for a time-dependent
nucleation rate.

In the snapshot showing nucleation, Fig.~\ref{fig:snapshot},
and in 
Supplementary Movie 1 \cite{supp_mat_pre14}, we see that nucleation occurs
with the nucleus in a concave part, or pit, in the surface.
This is as expected
\cite{page06,meel10,hedges12,page09_jacs,sear_rev12,sear_rev}.
The free energy barrier to nucleation comes from the cost
of creating the interface around the growing nucleus. The
shorter the length of interface that needs to be created,
the lower the free-energy barrier. At any surface
there is a pre-existing interface.
This pre-existing part of the interface does not contribute
to the barrier, which is why the barrier is lower at interfaces.
At any concave part of a surface, the length of this
pre-existing interface is
larger than at a flat or convex surface, and so nucleation is
faster there \cite{debenedetti,sear_rev,sear_rev12}.

At a supersaturation $2h/kT=0.12$, the radius of the critical
nucleus $r^*=\gamma/(2h)=8.75$. This used
Onsager's exact expression \cite{onsager44}, for the surface
tension $\gamma$, which gives
$\gamma=1.05 kT$ at $\epsilon/kT=3$.
This critical radius
$r^*$ is larger but of the same order
of magnitude as the roughness of the surface seen
in Fig.~\ref{fig:rough}.
A roughness of around 2, or about one quarter of the critical
radius is enough to dramatically increase the nucleation rate here.

The key finding of our computer simulations is that roughening
of the surface with time causes the nucleation rate to
increase (by orders of magnitude)
until it is fast enough to observe nucleation.
Although this is not an example of self-organised criticality (SOC)
\cite{bak88}, it does have features in common with
SOC systems. In both cases the dynamics drive
the system until a threshold is exceeded, at which point
a sudden event occurs (an avalanche in the sand pile model
of self-organised criticality, and nucleation here).
Future work could perhaps take ideas from the study
of the approach to the point where an avalanche occurs in SOC,
and apply them here.

\section{Analytical Model: Derivation, Results and Comparison with Computer Simulation
Results}

We will now develop a simple general model of
nucleation with a time-dependent rate. We do this in order to
generalise our findings for our lattice model, and to make
experimentally testable predictions.
We can construct a simple model for the increasing rate as follows.
The probability that nucleation has not occurred, $P(t)$, satisfies
the differential equation
\begin{equation}
\frac{{\rm d}P(t)}{{\rm d}t}=-P(t)r_N(t)
\label{eq:ode}
\end{equation}
where $r_N(t)$ is the nucleation rate of the system at time $t$.
It has dimensions of one over time.
Our observations require a rate that increases by orders of magnitude,
and so we assume that $r_N(t)$ is an exponentially increasing
function of time:
\begin{equation}
r_N(t)=r_0\exp\left[\lambda t\right]
\label{eq:exp_rate}
\end{equation}
where $r_0$ is the rate at $t=0$, and $\lambda$ is the rate
of increase of the nucleation rate.
This exponential time dependence will give us the required large increase in
rate, and would follow if the free-energy barrier is decreasing
linearly as a function of time.
The roughness is not increasing linearly with time,
see Fig.~\ref{fig:rough}, so this is unlikely to be exactly true,
but it does provide a simple model.
Note that with this assumption of an exponentially
increasing rate our model becomes the Gompertz model
\cite{mueller95}, a model widely used for predicting lifetimes,
for example the lifetimes of living organisms.

Putting Eq.~(\ref{eq:exp_rate}) into Eq.~(\ref{eq:ode}), and solving
we obtain \cite{mueller95}
\begin{eqnarray}
P(t)&=&\exp\left[(r_0/\lambda)\left(1-\exp[\lambda t]\right)\right]
\\
P(t)&\simeq&
\exp\left[-\exp\left[\lambda t+\ln\left(r_0/\lambda\right)\right]\right]
~~~\mbox{when}~~r_0/\lambda\ll 1~~~
\label{eq:exp_p}
\end{eqnarray}
We are interested in systems where the initial rate $r_0$
is much lower than $\lambda$, as then the rate increases
for some time before nucleation occurs. So, in the second line
we took the $r_0/\lambda\ll 1$ limit and simplified the equation.
It is worth noting that Eq.~(\ref{eq:exp_p}) has
the functional form of the Gumbel distribution of extreme-value
statistics \cite{castillo_book},
which can also arise in nucleation problems via other
mechanisms \cite{sear13_unpub}.

Fits of Eq.~(\ref{eq:exp_p}) to the simulation data are
shown as the brown and turquoise
dotted curves in Fig.~\ref{fig:bigp}. They
provide reasonably good fits to the data. The best fit values
of ($\lambda$,~$\ln(r_0/\lambda)$) are
($1.10\times 10^{-6}$,~$-3.75$) and ($1.77\times 10^{-6}$,~$-3.65$),
for the fits to data with dissolution rates
$r_D=10^{-6}$ and $2\times 10^{-6}$
respectively. 
So, the best-fit values
of $r_0/\lambda$ satisfy our assumption that $r_0/\lambda\ll 1$.

We can obtain an independent estimate of the value
of $\lambda$ by approximating it as the rate
of change of the free energy barrier to nucleation, $\Delta F^*$,
with time:
$\lambda\approx (\partial(\Delta F^*/kT)/\partial \rho)\times
(\partial \rho/\partial t)$, which is
expressed in terms of the rate of change of the roughness.
In Fig.~\ref{fig:rough} we see that the rate of increase
of $\rho$ is not a constant but is always of order $10^{-6}$
for $r_D=2\times 10^{-6}$, i.e., 
$(\partial \rho/\partial t)\approx r_D$.
We can estimate $(\partial(\Delta F^*/kT)/\partial \rho)$
by assuming that $\Delta F^*$ approximately
halves when $\rho$ becomes of order $r^*$.
At the supersaturation $2h/kT=0.12$, the initial nucleation
rate on a flat surface is
$1.76\times 10^{-8}\pm 0.74\times 10^{-8}$ (from FFS simulations).
Taking the barrier for a flat surface
$\Delta F^*/kT\approx \ln(r_0)\approx 18$. As $r^*\approx 8$,
we then have
$(\partial (\Delta F^*/kT)/\partial \rho)\approx (18/2)/8\approx 1$.
Thus, our estimate for $\lambda$ at $r_D=10^{-6}$ is
$\lambda\approx 10^{-6}$, which is close to the best
fit value.

Equation (\ref{eq:exp_p}) gives a standard deviation of nucleation times
$\sigma=(\pi/6^{1/2})/\lambda$ (obtaining
this expression requires extending the integration over $-\infty<t<\infty$
but for our parameter values this is a very good approximation).
Thus experimental
measurements of the spread of nucleation times
directly measure the timescale for the increase in nucleation rate.
The ratio of the standard deviation to the mean,
$\langle t_N\rangle$, is given by
$\sigma/\langle t_N\rangle=(\pi/6^{1/2})/(\ln(\lambda/r_0)-\gamma)$,
for $\gamma\simeq 0.577$ the Euler-Mascheroni constant.
In other words, for large $\lambda/r_0$,
$\sigma/\langle t_N\rangle$
scales as $1/\ln(\lambda/r_0)$, i.e.,
decreases but only as the log of the ratio. Thus
the prediction is that for systems with low initial rates,
$r_0$, the ratio $\sigma/\langle t_N\rangle$ should be significantly
less than one, but due to the logarithmic dependence, values
will presumably almost always be around $0.1$ or above.
Very small values are not achievable.

There is experimental data in which 
$\sigma / \langle t_N\rangle<1$, for example the work
of Fasano and Khan \cite{fasano01} on the crystallisation
from solution of calcium oxalate. A small value of
this ratio is characteristic of a system with a nucleation rate
increasing with time.
In the calcium oxalate
system, the surface the calcium oxalate crystals are nucleating
on may be changing with time in such a way as to increase the
rate of nucleation at this surface.

\section{Generality of our results and comparison with the initiation of cancer}

Our Eq.~(\ref{eq:exp_p}), by definition, will apply
to any process in which the nucleation rate varies exponentially
with time.
Our model of a slowly dissolving surface is just one member of a class
of systems, defined by having an increasing rate
of a stochastic process. In other systems, other processes may cause
the surface and hence the rate to change with time.
Another example of such a process may be
a chemical reaction that modified the surface such as to reduce the
contact angle of the nucleus at the surface.
Another might be nucleation on
some growing aggregate in solution, where the nucleation rate increases
as the size of the aggregate grows \cite{sear12}.
Nucleation of lysozyme crystals is known to be affected
by aggregates \cite{parmar07}.
All members of this class should have qualitatively similar
$P(t)$'s.

Finally, we note that our surfaces are an example of a
system where multiple random steps (not just one) are required before the event
of interest occurs. A number of erosion steps is needed, followed
by a nucleation step. Systems where multiple steps are required
may generically result in cumulative probabilities
$P(t)$ that are similar to those
in Fig.~\ref{fig:bigp}. An example in a very different context, but
with a $P(t)$ of a similar
form, is that of lung cancer.
The probability
that a smoker does not have lung cancer has a similar
form to our $P(t)$'s in Fig.~\ref{fig:bigp};
see Refs.~\onlinecite{halpern93,peto11}, where they plot
the probability of dying of cancer, equivalent to $1-P(t)$.
Smoking greatly increases the probability of getting lung cancer, but
for typical smokers who start when young,
this risk remains low until their 40s \cite{halpern93,peto11},
after which it rises rapidly.

\section{Conclusion}

Nucleation occurs on surfaces, and in practice all
surfaces change over time, it is simply a question of whether this
change is faster, slower, or comparable to the timescale of the experiment.
Almost all prior work studying nucleation has implicitly assumed
that it is much slower. 
Here we introduced a simple model
for nucleation on a slowly dissolving surface, and showed that it predicts
a characteristic $P(t)$ (Fig.~\ref{fig:bigp}).
This should be straightforward to observe in experiment.
It is also indicative of a mechanism determining
the time until nucleation, that is fundamentally
different to that predicted by classical nucleation theory.
Here, the time until nucleation
is the time taken for dissolution to
increase the nucleation rate to the point that it is fast enough
to occur before further change at the surface.
This is of course
qualitatively different from the mechanism in
classical nucleation theory, where the nucleation time
is the time for a rare thermal fluctuation
to occur.

A $P(t)$ consistent with a nucleation rate increasing with time
has been in observed by Kim {\it et al.}~\cite{kim13}.
Our model provides a possible qualitative explanation for this observation.
Toldy {\em et al.}~\cite{toldy12} also see a similar
$P(t)$, in that case for glycine crystallising in solution, but there the crystallising
droplets may not be independent of each other. They sometimes observe
that once one droplet has crystallised, then neighbouring droplets may
crystallise. This means droplets are not independent, which can give
an effective nucleation rate for the ensemble of droplets that increases
with time, even when the rate in an isolated droplet may not be increasing
with time. This effect complicates understanding the nucleation behaviour.

Further experimental studies will be needed to understand what sorts of 
crystallising systems have nucleation rates that vary with time, and to
understand what determines the key parameter: the rate at which the nucleation
rate increases with time, $\lambda$. If we are to predict and control
systems with time-dependent nucleation rates we will need to determine
what are the physical processes that control this time dependence.
Finally, computer simulations of crystallisation in simple off-lattice
models will be needed to better understand how crystals nucleate
on a time-dependent rough surface.

\begin{acknowledgments}
It is a pleasure to acknowledge discussions with Sathish Akella, Seth Fraden,
James Mithen, and Patrick Warren.
I acknowledge financial support from EPSRC (EP/J006106/1).
\end{acknowledgments}


%

\end{document}